# Single-crystal silver nanowires: Preparation and Surface-enhanced Raman Scattering (SERS) property

Baoliang Sun, Xiaohong Jiang, Shuxi Dai, Zuliang Du *

*Key Lab for Special Functional Materials of Ministry of Education, Henan University, Kaifeng 475004, PR China*

### ABSTRACT

Ordered Ag nanowire arrays with high aspect ratio and high density self-supporting Ag nanowire patterns were successfully prepared using potentiostatic electrodeposition within the confined nanochannels of a commercial porous anodic aluminium oxide (AAO) template. X-ray diffraction and selected area electron diffraction analysis show that the as-synthesized samples have preferred (220) orientation. Transmission electron microscopy and scanning electron microscopy investigation reveal that large-area and ordered Ag nanowire arrays with smooth surface and uniform diameter were synthesized. Surface-enhanced Raman Scattering (SERS) spectra show that the Ag nanowire arrays as substrates have high SERS activity.

*Keywords:*
Silver nanowire
AAO
SERS

## 1. Introduction

As one of the most important one-dimensional nanostructures, metal nanowires have attracted a great deal of research interest in recent years because of their electronic and optical properties [1,2] and potential applications in nanodevices [3,4]. Among them, silver nanowires or nanorods have been recently prepared and studied by several groups using templates [1–8]. Generally, the template method is cheap and easy to operate, and can produce nanowires with uniform diameters ranging from several nanometers to hundreds of nanometers in large area. Possible templates include nuclear track-etched polycarbonate membranes, nanochannel glasses and anodic aluminum oxide (AAO) templates. It has also been found that the AAO template is an ideal template because it has many desirable characteristics, including tunable pore dimensions over a wide range of diameters and lengths, good mechanical strength and thermal stability. Among the well-developed fabrication methods based on AAO template, electrodeposition is recognized as one of the most efficient methods in controlling the growth of nanowires and has been widely used to produce a variety of metal [9,10], semiconductor, and conductive polymer nanowire arrays.

On the other hand, Surface-enhanced Raman Scattering (SERS) is a routine and powerful tool for the investigation and structural characterization of interfacial and thin-film systems [1]. Because the surface morphologies and structures of the substrates determine the generation and intensity of Raman signals, the substrates play a vital role in SERS and the research on the SERS active substrates remains a hot topic. Usually the metal (especially silver) island membrane, roughened metal electrodes, silver sol, chemical etching and chemical deposition Ag membrane and so on were found to be appropriate for SERS substrates[11]. Experimentally, the challenge of achieving an ideal reproducible surface morphology on a metal surface can be quite daunting. This fundamental limitation demonstrates the need for fabrication methods that produce reproducible and practical SERS-active substrates. As we known, there is very few paper reported on Ag nanowire arrays as SERS substrate. So, the preparation of long and continuous silver nanowires (with high aspect ratio, and with well controlled morphology and microstructures) still remains its importance.

In this work, we report the electrochemical fabrication of single-crystalline silver nanowire arrays in commercially available AAO templates with a nominal pore diameter of 100 nm. The silver nanowire arrays were appropriate for SERS substrates using pyridine as a reported molecule.

## 2. Experiment

Ag nanowire arrays were synthesized using an AAO template with a nominal pore diameter of 100 nm (Whatman Company) at room temperature in air. The membranes typically had a thickness of 60 μm. Before electrodeposition, a thin gold layer (about 200 nm) was evaporated onto one side of an AAO template, which served as the working electrode in the subsequent electrodeposition process. The electrochemical deposition solution contains 18 g l$^{-1}$ AgNO$_3$ (analytical reagent, AR), 120 g l$^{-1}$ trisodium citrate dehydrate (C$_6$H$_5$Na$_3$O$_7$·2H$_2$O, AR), 40 g l$^{-1}$ potassium sodium tartrate tetrahydrate (C$_4$H$_4$O$_6$KNa·4H$_2$O, AR), 60 g l$^{-1}$ sodium sulfite (Na$_2$SO$_3$, AR). The pH value of the solution was adjusted to the range of 6 by adding the H$_3$BO$_3$ and NH$_3$·H$_2$O solutions. In the preparation process, the deposition potential was firstly kept at 1 V for 60 s to make the silver nucleate at the bottom of the pores, then the deposition current was adjusted to 0.5 mA cm$^{-2}$ for 40h by decreasing the voltage.

* Corresponding author. Tel./fax: +86 378 3881358.
  E-mail address: zld@henu.edu.cn (Z. Du).

Structural characterization of the silver nanowires embedded in alumina membranes was examined by X-ray diffraction (XRD, X'pert MPD-Philips diffractometer with Cu K$_\alpha$ radiation, $\lambda = 0.15418$ nm, Philips, Holland). The morphology and structure of the samples were observed by scanning electron microscopy (SEM, JSM-5600, JEOL, Japan). Transmission electron microscope ((TEM) JEM-100CX TEM operating at 100 kV) and selected-area electron diffraction (SAED) were employed to characterize the individual Ag nanowires. The Raman and SERS spectra were collected using a Renishaw RM-1000 co focal microscopic Raman spectrometer with an air-cooled CCD detector, the operating wavelength was 632.8 nm by He–Ne laser and the output laser power was 1.3 mW at the sample stage.

## 3. Results and discussion

### 3.1. Silver nanowire growth mechanism

Silver electrodeposition can be carried out in aqueous solution from its solvated ions or from its hydro soluble complexes. The disadvantage of silver electrodeposition from AgNO$_3$ solutions is the poor quality of the electroformed metal [2]. Thus, the electrodeposition from silver complex solutions is preferred. Previous research shows that the deposition of silver nanowires from solutions formed principally with a diversity of additives: acetaldehyde [12], cyanide [10], potassium thiocyanate [8] and nitric acid [2]. Some of these solutions are toxicity, or the pH values are too low or too high, so they are not applicable for AAO template because the solubility process of amorphous alumina membranes begins at pH<5 and pH>8.2 [13]. Therefore, it was necessary to find a ligand for Ag(I) to be stable at neutral and slightly acidic pHs.

In this work we found that using chemical reagent of C$_6$H$_5$Na$_3$O$_7$·2H$_2$O and C$_4$H$_4$O$_6$KNa·4H$_2$O as ligands together have good effects to synthesis silver nanowires with single-crystal structure. In addition, this route avoids the use of the highly poisonous cyanide solutions, which was currently used in electrochemical deposition of noble metals and it can work at pH>4 and pH<8. It should also be mentioned that the electrodeposition is a complicated process which involves charge transfer, diffusion, reaction, adsorption on substrate. Therefore, the structure of silver nanowires is closely related to deposition conditions and its growth modes, the critical dimension $Nc$ is inverse proportional to the overpotential square [9].

One possible reason for the single-crystal growth of nanowires is the 2D-like nucleus under lower overpotential, because the smaller overpotential, the larger $Nc$, then the more favorable for the single-crystal growth of nanowires. So we choose low deposition voltage to get single-crystal Ag nanowires.

### 3.2. Structural and morphological studies of the silver nanowire arrays

Fig. 1 shows the typical XRD pattern of the as-synthesized Ag nanowire arrays when the deposition potential was kept at 0.2 V for 40h. The diffraction peaks are corresponding to (111) and (220) reflections of the face-centered cubic (fcc) structure of bulk Ag with the standard lattice constant of $a = 4.077°$A (JCPDS 87-0720), respectively. It is obviously that the intensity of peak at 64.46° is much stronger than that at 38.18° (111), which means that our silver nanowires have preferred (220) orientation. In fact, for Ag, the close-packed (111) face has the lowest energy, followed by the (100) and (110) faces [9]. For lower overpotentials, it is well known that the (111) face is preferred due to its low surface energy for the thermodynamic reason. At this point the influence of competition between the adsorption and desorption of H ions at the frontier growth surfaces of metallic nanowires must be taken into account.

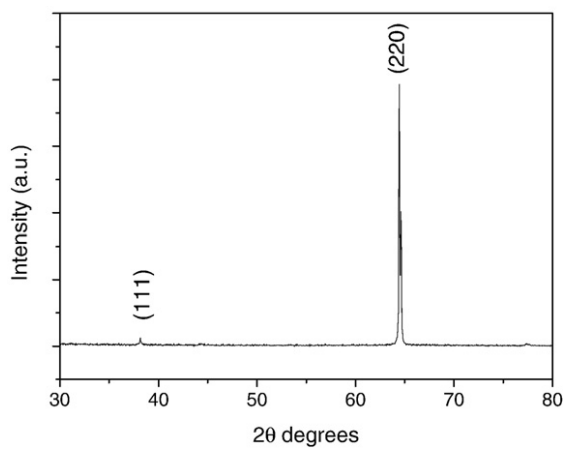

**Fig. 1.** X-ray diffraction pattern of the silver nanowire arrays embedded in the alumina template. The deposition potential was kept at 0.2 V for 40h.

Slightly higher overpotentials kinetically favor the formation of (220) and the H adsorption stabilizes the process [2]. Then, change of the overpotential induced the thermodynamic to kinetic transition and favored the (220) face at slightly higher overpotentials, stabilized by means of the H ion adsorption, resulting in the preferred (220) orientation observed in the metallic crystallographic Ag nanowires.

Fig. 2 shows the SEM images of the as-synthesized samples after removing the top alumina layer with 3 M NaOH illustrate that the products are well-aligned nanowires. Fig. 2(a)–(b) are top views of the sample and Fig. 2(c) is a side view of the arrays. We can see that the length of nanowires was only about 20 μm in average at 0.2 V above 40h, which was shorter than the (AAO) templates thickness (about 60 μm). This indicated that the growth speed of nanowires was very slow at this depositing condition and if deposition time was increased or decreased, longer or shorter nanowires could be obtained, respectively. Fig. 2(d) shows the SEM image for the Ag nanowires arrays when the deposition potential was kept at 0.2 V for 3 h. In this case the length of nanowires was only about 2 μm. On the other hand, if the deposition bias was increased, the growth speed of nanowires could quicken obviously and long nanowires could be obtained at short deposition time. In this way, we can easily control the height of Ag nanowire arrays by deposition time and bias.

One single Ag nanowire structure and morphology were further studied with TEM (Fig. 3). Fig. 3(a) shows that one Ag nanowire is very straight with diameter of about 100 nm, which is nearly equal to the channel diameter of the AAO template. The SAED pattern taken from this nanowire corresponds to Ag (220) as shown in Fig. 3(b). The SAED pattern indicates that the nanowires are high-quality single crystal, which is consistent with discussion in Section 3.1. Fig. 3(c) shows the top of Ag nanowire TEM image. One can see the top of nanowire is very smooth, which should result from slowly deposition.

### 3.3. SERS of pyridine used silver nanowire arrays as substrates

In spite of SERS being very popular, it does have many limitations, including how to achieve optimal enhancement. One of the critical aspects of this technique involves the need for producing an ideal surface morphology on SERS substrate for maximum enhancement, which is predicted from long-range classical electromagnetic (EM) theory [14].

Here Ag nanowire arrays shown as in Fig. 2(d) with about 2μm in lengths were used as SERS-active substrate and pyridine was used as the measured molecule. Typical SERS spectra are presented in Fig. 4(a). The SERS surface enhancement factor (SEF) [1] was calculated for pyridine on Ag nanowire arrays according to the equation $SEF = (I_{surf}/N_{surf})/(I_{bulk}/N_{bulk})$. In this expression, $I_{surf}$ and $I_{bulk}$ denote the integrated intensities for the 1007 cm$^{-1}$ band of the pyridine adsorbed on Ag surface and pyridine in solution, respectively; whereas $N_{surf}$ and $N_{bulk}$ represent the corresponding number of pyridine molecules excited by laser beam. The maximum average SERS SEF calculated for multiple locations on these Ag nanowires was $2.2 \times 10^6$, which is two orders of magnitude larger than those obtained from previously published methods of forming nanoparticle arrays by vapor deposition SEF $10^4$ [15,16]. It is possible that the large SERS enhancement factors for the substrate are mainly due to the long-range classical electromagnetic dipole effect accentuated by the high aspect ratio of the Ag nanowire arrays, which is similar to the SERS of Ag nanorods [1]. So, the Ag nanowire arrays can be applied as high sensitivity substrates for SERS-based measurements, which were fabricated utilizing electrodeposition technique implemented easily in the laboratory.

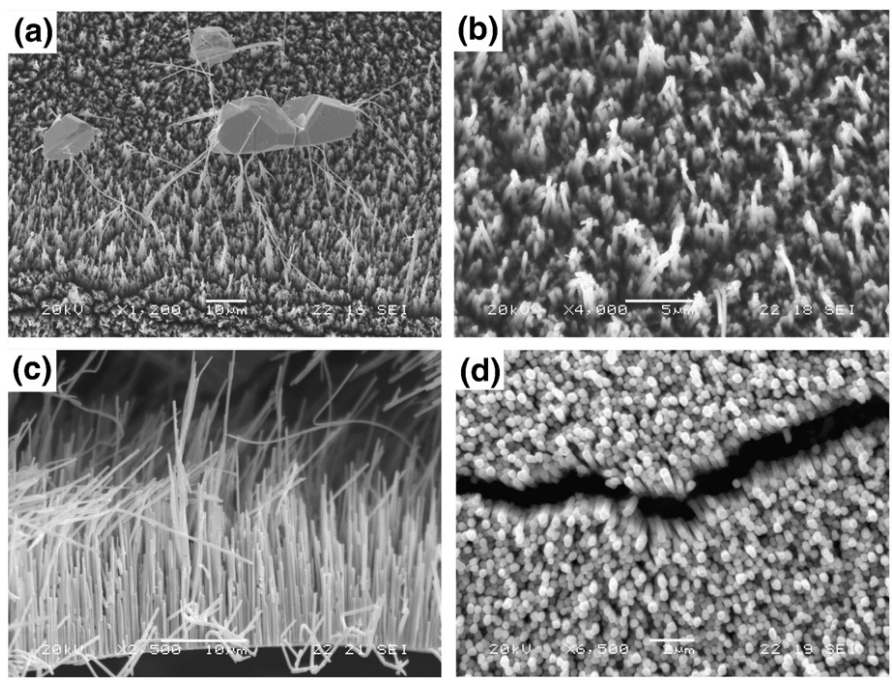

**Fig. 2.** SEM images of the silver nanowires arrays after dissolving the alumina membrane when the deposition potential was kept at 0.2 V above 40 h. (a) Top view of the sample after dissolving the alumina membrane. (b) Enlarged image of (a). (c) Side view of the sample. (d) SEM image of shorter Ag nanowires arrays when the deposition potential was kept at 0.2 V for 3 h.

## 4. Conclusion

In summary, we have prepared the single-crystal silver nanowire arrays by utilizing AAO template combined with electrodeposition. The height of Ag nanowire arrays can be easily controlled from two micrometers to twenty micrometers by varying deposition time and bias, while the diameter of nanowire keeps invariable (~100 nm). These arrays show potential as high sensitivity substrates for SERS-based measurements. For the arrays of Ag nanowires with a length of 2 μm the SERS enhancement factor can achieve above $10^6$.

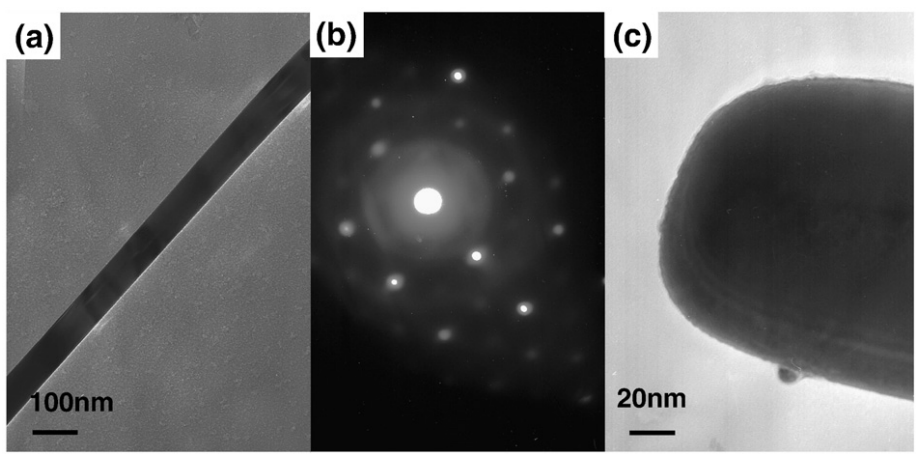

**Fig. 3.** (a) TEM image of one single silver nanowire with about 100 nm diameter; (b) SAED pattern from the nanowire in (a) indicates the single-crystalline Ag nanowire has preferred (220) orientation; (c) TEM image of the top silver nanowire.

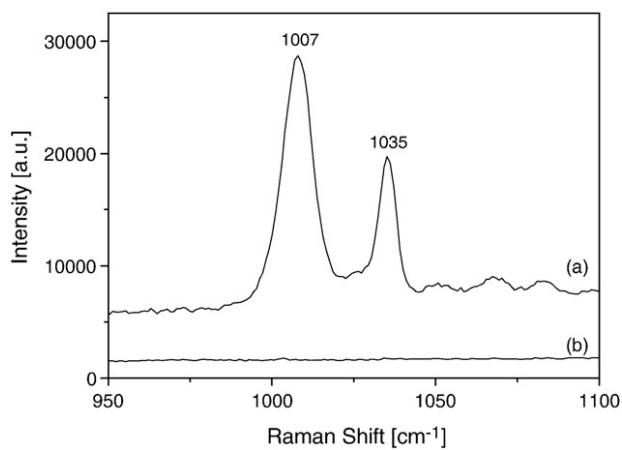

**Fig. 4.** (a) The typical SERS spectra of pyridine adsorbed onto Ag nanowire arrays with 2 μm in length. (b) The SERS spectra of pyridine in solution with 0.01 M.

### Acknowledgment

This work was supported by the Natural Science Foundation of China (Grant Nos. 90306010, 10874040 and 20803018), State Key Basic Research "973" Plan of China (No. 2007CB616911), Program for New Century Excellent Talents in University of China (NCET-04-0653) and the Cultivation Fund of the Key Scientific and Technical Innovation Project, Ministry of Education of China (No. 708062).